\documentclass[12pt,a4paper]{article}
\usepackage{amsmath,amssymb}
\usepackage{graphicx}

\allowdisplaybreaks[3]

\begin{document}

\begin{titlepage}

\begin{flushright}
KEK-TH-1976, KUNS-2674
\end{flushright}

\vspace{5em}

\begin{center}
{\Large\bf Non-Gaussian and  loop effects of inflationary correlation functions in BRST formalism}
\end{center}

\begin{center}
Hiroyuki K{\sc itamoto}$^{1)}$
\footnote{E-mail address: kitamoto@tap.scphys.kyoto-u.ac.jp}
Yoshihisa K{\sc itazawa}$^{2),3)}$
\footnote{E-mail address: kitazawa@post.kek.jp} 
Ryota K{\sc ojima}$^{3)}$
\footnote{E-mail address: ryota@post.kek.jp} 

\end{center}

\begin{center}
$^{1)}$
{\it Division of Physics and Astronomy}\\
{\it Graduate School of Science, Kyoto University}\\
{\it Kyoto 606-8502, Japan}\\
$^{2)}$
{\it KEK Theory Center, Tsukuba, Ibaraki 305-0801, Japan}\\
$^{3)}$
{\it Department of Particle and Nuclear Physics}\\
{\it The Graduate University for Advanced Studies (Sokendai)}\\
{\it Tsukuba, Ibaraki 305-0801, Japan}
\end{center}

\begin{abstract}
We investigate inflationary correlation functions in single field inflation models. 
We adopt a BRST formalism where locality and covariance at the sub-horizon scale are manifest. 
The scalar and tensor perturbations are identified with those in the comoving gauge which become constant outside the cosmological horizon. 
Our construction reproduces the identical non-Gaussianity with the standard comoving gauge. 
The accumulation of almost scale invariant fluctuations could give rise to  IR logarithmic corrections at the loop level. 
We investigate the influence of this effect on the sub-horizon dynamics.
Since such an effect must respect covariance,
our BRST gauge has an advantage over the standard comoving gauge. 
We estimate IR logarithmic effects to the slow-roll parameters at the one-loop level. 
We show that $\epsilon$ receives IR logarithmic corrections, while this is not the case for $\eta$. 
We point out that IR logarithmic effects provide the shift symmetry breaking mechanism.
This scenario may lead to an inflation model with a linear potential.

\end{abstract}

\vspace{\fill}

July 2017

\end{titlepage}

\section{Introduction}
\setcounter{equation}{0}

The central dogma of inflation theory is that the cosmological perturbations (scalar and tensor modes) are frozen at the super-horizon scale.  
These fluctuations are almost scale invariant \cite{Mukhanov,Starobinsky,Hawking,Guth,Bardeen} and so they could give rise to IR logarithmic corrections to the correlation functions through loop effects as 
\begin{align}
\int_{k<Ha(t)}\ \frac{d^3k}{k^3}\ \sim \log a(t). 
\end{align}

The IR logarithmic corrections to the cosmological perturbations cancel at the super-horizon scale up to the leading order of slow-roll. 
This fact can be demonstrated by the following argument. 
The scalar $\zeta$ and tensor $\gamma_{ij}$ perturbations in  the comoving gauge are 
\begin{align}
g_{ij}=a^2(t)e^{2\zeta}(e^{\gamma})_{ij},\hspace{1em}\hat{\varphi}, 
\end{align}
where $g_{ij}$ is the spatial metric and $\hat{\varphi}$ denotes the classical solution of the inflaton.
In ADM formalism, the lapse and shift variables are determined by the equation of motion, as briefly reviewed in Appendix \ref{A}. 
The quadratic action for the scalar perturbation in the comoving gauge is
\begin{align}
\int d^4x\ \frac{2\epsilon}{\kappa^2}\big[a^3\dot{\zeta}\dot{\zeta}-a\partial_i\zeta\partial_i\zeta\big]. 
\label{suphor}
\end{align}
There is no potential term for the scalar perturbation, since the constant shift $\zeta \rightarrow  \zeta+c$ can be canceled by rescaling the spatial coordinates. 
The kinetic term is proportional to the slow-roll parameter $\epsilon$ as it becomes a pure gauge in the de Sitter limit. 
Therefore, the scalar perturbation becomes constant outside the horizon in the comoving gauge. 

Another gauge which is suited to investigate sub-horizon physics is
\begin{align}
g_{ij}=a^2(t)(e^{\gamma})_{ij},\hspace{1em}\hat{\varphi}+\varphi. 
\end{align}
The quadratic action for the scalar field $\varphi$ is
\begin{align}
\int d^4x\ \frac{1}{2}\big[a^3\dot{\varphi}\dot{\varphi}-a\partial_i\varphi\partial_i\varphi -(3\eta-6\epsilon)H^2a^3\varphi^2\big]. 
\label{subhor}
\end{align}
They are related by the gauge transformation: 
\begin{align}
\zeta=-H\delta t=-\frac{\kappa\varphi}{2\sqrt{\epsilon}},\hspace{1em}\delta t =\frac{\varphi}{\dot{\hat{\varphi}}}. 
\label{gaugetr}
\end{align}
The evolution of the scalar field $\varphi$ can be traced by the action (\ref{subhor}) up to the horizon scale. 
The action (\ref{suphor}) shows that the scalar perturbation $\zeta$ becomes constant outside the horizon after the gauge transformation (\ref{gaugetr}). 
In fact, the time dependence of $\dot{\epsilon}/\epsilon=2H(2\epsilon-\eta)$
is precisely canceled by the time dependence of $\dot{{\varphi}}/\varphi=-m^2/3H$
as it possesses the consistent mass $m^2=3(\eta-2\epsilon)H^2$ in (\ref{subhor}).

In this way, the scalar perturbation is estimated as
\begin{align}
\langle\zeta_{\bf k} (t)\zeta_{{\bf k}'}(t)\rangle=\frac{\kappa^2}{4\epsilon_*}\langle\varphi_{\bf k}(t)\varphi_{{\bf k}'}(t)\rangle_*
=(2\pi)^3\delta^{(3)} ({\bf k}+{\bf k}')\cdot\frac{\kappa^2}{4\epsilon_*}\frac{H_*^2}{2k^3}, 
\end{align}
where the star  means it is evaluated at the time of horizon crossing.
Although we retain only physical modes (scalar and tensor) in these gauges, the price to pay is that they are highly non-local.
They are analogues of the Coulomb gauge in QED. 

It is possible to adopt a local gauge instead, as explained in Section 2. 
We parametrize the metric as $g_{\mu\nu}=a^2(t)e^{2\kappa\omega}\tilde{g}_{\mu\nu}=a^2(t)e^{2\kappa\omega}(e^{\kappa h})_{\mu\nu}$.
We adopt a BRST gauge where the spin-$0$  mode $h_{00}=2\omega, h_{00}=h^i_{~i}$ and the traceless spin-$2$ mode $\tilde{h}_{ij}$ are massless and minimally coupled.
The inflaton $\varphi$ couples minimally to gravity, 
\begin{align}
\int d^4x\ \frac{1}{2}a^2\big[e^{2\kappa h_{00}}\partial_0\varphi\partial_0\varphi-e^{\frac{2}{3}\kappa h_{00}}\tilde{g}^{ij}\partial_i\varphi\partial_j\varphi\big], 
\end{align}
where the small mass term is suppressed. 
We consider the IR logarithmic effect to the scalar perturbation due to the $h_{00}$ and $\tilde{h}_{ij}$ modes. 
We may rescale $\varphi\to e^{-\kappa h_{00}}\varphi$ since we can ignore derivatives of $h_{00}$ 
\begin{align}
\int d^4x\ \frac{1}{2}a^2\big[\partial_0\varphi\partial_0\varphi-e^{-\frac{4}{3}\kappa h_{00}}\tilde{g}^{ij}\partial_i\varphi\partial_j\varphi\big]. 
\end{align}

The argument can be given for the cancellation of IR logarithmic effects to the scalar perturbation at the super-horizon scale \cite{Giddings}. 
First, with respect to the spin-$0$ mode  $h_{00}$, 
\begin{align}
\varphi^2\to e^{-2\kappa h_{00}}\varphi^2\sim e^{-2\kappa h_{00}}(e^{-\frac{4}{3}\kappa h_{00}}\tilde{g}^{ij}k_ik_j)^{-\frac{3}{2}} \sim (\tilde{g}^{ij}k_ik_j)^{-\frac{3}{2}}. 
\end{align}
Second, with respect to the spin-$2$ mode, 
\begin{align}
\langle(\tilde{g}^{ij}k_ik_j)^{-\frac{3}{2}} \rangle\sim k^{-3}. 
\end{align}

On the other hand, IR logarithmic effects do not cancel at the sub-horizon scale.
In nonlinear sigma models, the coupling evolves with time due to IR logarithmic effects \cite{Davis,KitamotoNS}.
The action (\ref{suphor}) is reminiscent of nonlinear sigma models where $\epsilon$ plays the role of the inverse coupling. 
We have argued that they make the couplings of a generic  theory time dependent in de Sitter spacetime
\cite{KitamotoSD,KitamotoG}. 

Our strategy is to investigate IR logarithmic effects for sub-horizon dynamics. 
Cosmological perturbations originate from the quantum fluctuation at the sub-horizon scale. 
We can estimate their magnitude at horizon crossing---namely, at the boundary
of the sub-horizon effective theory. 
The slow-roll inflation theory is characterized by two small parameters $\epsilon$ and $\eta$. 
They represent the local slope and curvature of the inflaton potential. 
They evolve with time even at classical revel. 
We show that $\epsilon$ receives IR logarithmic effects which give rise to quantum time evolution beyond the classical one. 
On the other hand, $\eta$ does not receive IR logarithmic corrections. 

One of the major puzzles of inflation theory is to explain why slow-roll parameters are small.
Shift symmetry with respect to the inflaton field $\varphi\rightarrow \varphi +c$ may be necessary to explain it.  
It is often broken by hand to construct inflation models, leading to a proliferation of
them. We need a better understanding of this symmetry breaking mechanism.
The IR effect to $\epsilon$ could solve this difficulty as this quantum effect may provide a shift symmetry breaking mechanism.
This scenario may lead to an inflation model with a linear potential.

In Section 2, we introduce the propagators in BRST formalism. 
The advantage of this gauge over the standard comoving gauge is its locality. 
The covariance at the sub-horizon scale is more manifest in this gauge. 
These features allow us to identify IR logarithmic effects at the sub-horizon scale. 
In Section 3, we investigate the non-Gaussianity of the cosmological correlators. 
We explain how to reproduce the cosmological correlators which are identical with those in the comoving gauge and become constant outside the horizon. 
In Section 4 we investigate infrared logarithmic effects  to scalar perturbation. 
We show that $\epsilon$ receives nontrivial correction to the leading order of slow-roll parameters, 
while this is not the case for $\eta$. 
We conclude in Section 5. 

\section{Propagators in BRST formalism}
\setcounter{equation}{0}

We consider the action of a single field slow-roll inflation: 
\begin{align}
S=\int\sqrt{-g}d^4x\ \big[\frac{M_\text{pl}^2}{2}R-\frac{1}{2}g^{\mu\nu}\partial_\mu(\hat{\varphi}+\varphi)\partial_\nu(\hat{\varphi}+\varphi)-V(\hat{\varphi}+\varphi)\big]. 
\label{action}\end{align}

Assuming that the background is homogeneous, isotropic and spatially flat 
\begin{align}
\hat{g}_{\mu\nu}dx^\mu dx^\nu&=-dt^2+a^2(t)d{\bf x}^2 \notag\\
&=a^2(\tau)(-d\tau^2+d{\bf x}^2), 
\end{align}
the background fields $a, \hat{\varphi}$ satisfy the following equations:  
\begin{align}
3M_\text{pl}^2H^2=\frac{1}{2}\dot{\hat{\varphi}}^2+\hat{V}, 
\end{align}
\begin{align}
M_\text{pl}^2\dot{H}=-\frac{1}{2}\dot{\hat{\varphi}}^2, 
\end{align}
\begin{align}
\ddot{\hat{\varphi}}+3H\dot{\hat{\varphi}}+\hat{V}'=0, 
\end{align}
where $H=\dot{a}/a$, $\dot{a}=\partial_t a$ and $V'=\partial V/\partial \varphi$. 

We parametrize the metric fluctuation as 
\begin{align}
g_{\mu\nu}=\Omega^2\tilde{g}_{\mu\nu},\hspace{1em}\Omega=ae^{\kappa \omega}, 
\end{align}
\begin{align}
\tilde{g}_{\mu\nu}=\eta_{\mu\rho}(e^{\kappa h})^\rho_{\ \nu},\hspace{1em}h_\mu^{\ \mu}=0, 
\end{align}
where $\kappa\equiv \sqrt{2}/M_\text{pl}$. 
In the parametrization, the action (\ref{action}) is written as 
\begin{align}
S=\int d^4x\ \big[&\frac{1}{\kappa^2}(\Omega^2\tilde{R}+6\tilde{g}^{\mu\nu}\partial_\mu\Omega\partial_\nu\Omega) \notag\\
&-\frac{1}{2}\Omega^2\tilde{g}^{\mu\nu}\partial_\mu(\hat{\varphi}+\varphi)\partial_\nu(\hat{\varphi}+\varphi)-\Omega^4V(\hat{\varphi}+\varphi)\big], 
\label{action'}\end{align}
where $\tilde{R}$ is the Ricci scalar of $\tilde{g}_{\mu\nu}$: 
\begin{align}
\tilde{R}=-\partial_\mu\partial_\nu \tilde{g}^{\mu\nu}-\frac{1}{4}\tilde{g}^{\mu\nu}\tilde{g}^{\rho\sigma}\tilde{g}^{\alpha\beta}\partial_\mu\tilde{g}_{\rho\alpha}\partial_\nu\tilde{g}_{\sigma\beta}
+\frac{1}{2}\tilde{g}^{\mu\nu}\tilde{g}^{\rho\sigma}\tilde{g}^{\alpha\beta}\partial_\mu\tilde{g}_{\sigma\alpha}\partial_\rho\tilde{g}_{\nu\beta}. 
\end{align}

The quadratic terms of the Lagrangian are given by 
\begin{align}
\mathcal{L}_2=&-\frac{1}{4}a^2\partial_\mu h_{\rho\sigma}\partial^\mu h^{\rho\sigma}+(3H^2-\frac{1}{4}\kappa^2\dot{\hat{\varphi}}^2)a^4h^0_{\ \rho}h^{0\rho} \notag\\
&+6a^2\partial_\mu \omega\partial^\mu \omega -8(3H^2-\frac{1}{4}\kappa^2\dot{\hat{\varphi}}^2)a^4\omega^2 \notag\\
&-\frac{1}{2}a^2\partial_\mu\varphi\partial^\mu\varphi-\frac{1}{2}\hat{V}''a^4\varphi^2-4\kappa\hat{V}'a^4\omega\varphi \notag\\
&+\frac{1}{2}a^2\partial_\mu h^{\mu}_{\ \rho}\partial_\nu h^{\nu\rho}+2Ha^3h^0_{\ \rho}\partial_\mu h^{\mu\rho}-2a^2\partial_\mu h^{\mu\nu}\partial_\nu \omega-8Ha^3h^{0\mu}\partial_\mu \omega \notag\\
&+\kappa\dot{\hat{\varphi}}a^3h^{0\mu}\partial_\mu\varphi+2\kappa\dot{\hat{\varphi}}a^3\omega\partial_0\varphi. 
\label{quadratic}\end{align}

In order to fix the gauge degrees of freedom, we introduce the following term into the Lagrangian \cite{Woodard1998,Woodard2001}: 
\begin{align}
\mathcal{L}_\text{GF}&=-\frac{1}{2}a^2F_\mu F^\mu, \notag\\
F_\mu&=\partial_\rho h^\rho_{\ \mu}-2\partial_\mu \omega+2Ha h_\mu^{\ 0}+4Ha\delta_\mu^{\ 0}\omega-\kappa\dot{\hat{\varphi}}a\delta_\mu^{\ 0}\varphi. 
\label{GF}\end{align}
After partial integrations, the gauge fixing term is written as 
\begin{align}
\mathcal{L}_\text{GF}=&-2H^2a^4h^0_{\ \rho}h^{0\rho}-2a^2\partial_\mu \omega\partial^\mu \omega+(20H^2-\kappa^2\dot{\hat{\varphi}}^2)a^4\omega^2+\frac{1}{2}\kappa^2\dot{\hat{\varphi}}^2a^4\varphi^2 \notag\\
&+(4H^2-\kappa^2\dot{\hat{\varphi}}^2)a^4h^{00}\omega-(\kappa H\dot{\hat{\varphi}}+\kappa\ddot{\hat{\varphi}})a^4h^{00}\varphi-(10\kappa H\dot{\hat{\varphi}}+2\kappa\ddot{\hat{\varphi}})a^4\omega\varphi \notag\\
&-\frac{1}{2}a^2\partial_\mu h^{\mu}_{\ \rho}\partial_\nu h^{\nu\rho}-2Ha^3h^0_{\ \rho}\partial_\mu h^{\mu\rho}+2a^2\partial_\mu h^{\mu\nu}\partial_\nu \omega+8Ha^3h^{0\mu}\partial_\mu \omega \notag\\
&-\kappa\dot{\hat{\varphi}}a^3h^{0\mu}\partial_\mu\varphi-2\kappa\dot{\hat{\varphi}}a^3\omega\partial_0\varphi. 
\label{GF'}\end{align}
The sum of (\ref{quadratic}) and (\ref{GF'}) is given by 
\begin{align}
\mathcal{L}_2+\mathcal{L}_\text{GF}=&-\frac{1}{4}a^2\partial_\mu \tilde{h}^{ij}\partial^\mu \tilde{h}^{ij}+\frac{1}{2}a^2\partial_\mu h^{0i}\partial^\mu h^{0i}+(H^2-\frac{1}{4}\kappa^2\dot{\hat{\varphi}}^2)a^4h^{0i}h^{0i} \notag\\
&-\frac{1}{3}a^2\partial_\mu h^{00}\partial^\mu h^{00}-(H^2-\frac{1}{4}\kappa^2\dot{\hat{\varphi}}^2)a^4h^{00}h^{00} \notag\\
&+4a^2\partial_\mu \omega\partial^\mu \omega -4(H^2-\frac{1}{4}\kappa^2\dot{\hat{\varphi}}^2)a^4\omega^2 \notag\\
&-\frac{1}{2}a^2\partial_\mu\varphi\partial^\mu\varphi-\frac{1}{2}(\hat{V}''-\kappa^2\dot{\hat{\varphi}}^2)a^4\varphi^2 \notag\\
&+4(H^2-\frac{1}{4}\kappa^2\dot{\hat{\varphi}}^2)a^4h^{00}\omega-(\kappa H\dot{\hat{\varphi}}+\kappa\ddot{\hat{\varphi}})a^4h^{00}\varphi \notag\\
&-(10\kappa H\dot{\hat{\varphi}} +2\kappa\ddot{\hat{\varphi}}+4\kappa\hat{V}')a^4\omega\varphi, 
\label{sum}\end{align}
where the spatial metric is decomposed into the traceless part and the trace part: 
\begin{align}
h^{ij}=\tilde{h}^{ij}+\frac{1}{3}\delta^{ij}h^{00}. 
\end{align}

Imposing the slow-roll condition: 
\begin{align}
M_\text{pl}^2H^2\gg \dot{\hat{\varphi}}^2,\hspace{1em}\ddot{\hat{\varphi}}\gg H\dot{\hat{\varphi}}, 
\end{align}
we introduce the slow-roll parameters: 
\begin{align}
\epsilon\equiv \frac{1}{2}\big(\frac{M_\text{pl}\hat{V}'}{\hat{V}}\big)^2\sim \frac{1}{2}\frac{\dot{\hat{\varphi}}^2}{M_\text{pl}^2H^2}\ll 1, 
\end{align}
\begin{align}
\eta\equiv \frac{M_\text{pl}^2\hat{V}''}{\hat{V}}\sim-\frac{\ddot{\hat{\varphi}}}{H\dot{\hat{\varphi}}}+\frac{1}{2}\frac{\dot{\hat{\varphi}}^2}{M_\text{pl}^2H^2}\ll 1. 
\end{align}
Sometimes it is useful to express $\eta$ by $\epsilon$: 
\begin{align}
\eta=2\epsilon-\frac{\dot{\epsilon}}{2H\epsilon}. 
\label{classical}\end{align}

Up to the leading order of the slow-roll parameters, (\ref{sum}) is written as 
\begin{align}
\mathcal{L}_2+\mathcal{L}_\text{GF}=&-\frac{1}{4}a^2\partial_\mu \tilde{h}^{ij}\partial^\mu \tilde{h}^{ij}+\frac{1}{2}a^2\partial_\mu h^{0i}\partial^\mu h^{0i}+H^2a^4h^{0i}h^{0i} \notag\\
&-\frac{1}{3}a^2\partial_\mu h^{00}\partial^\mu h^{00}-H^2a^4h^{00}h^{00} \notag\\
&+4a^2\partial_\mu \omega\partial^\mu \omega -4H^2a^4\omega^2 \notag\\
&-\frac{1}{2}a^2\partial_\mu\varphi\partial^\mu\varphi-\frac{1}{2}(3\eta-4\epsilon) H^2a^4\varphi^2 \notag\\
&+4H^2a^4h^{00}\omega-2\sqrt{\epsilon}H^2a^4h^{00}\varphi+4\sqrt{\epsilon}H^2a^4\omega\varphi \notag\\
=&-\frac{1}{4}a^2\partial_\mu \tilde{h}^{ij}\partial^\mu \tilde{h}^{ij}+\frac{1}{2}a^2\partial_\mu h^{0i}\partial^\mu h^{0i}+H^2a^4h^{0i}h^{0i} \notag\\
&+\frac{1}{2}a^2\partial_\mu X\partial^\mu X-\frac{1}{2}a^2\partial_\mu Y\partial^\mu Y-H^2a^4Y^2 \notag\\
&-\frac{1}{2}a^2\partial_\mu \Phi\partial^\mu\Phi-\frac{1}{2}(3\eta-6\epsilon)H^2a^4\Phi^2, 
\end{align}
where $X$, $Y$ and $\Phi$ are defined as 
\begin{align}
X&\equiv -\frac{1}{\sqrt{3}}h^{00}+2\sqrt{3}\omega, \notag\\
Y&\equiv h^{00}-2\omega+\sqrt{\epsilon}\varphi, \notag\\
\Phi&\equiv -\sqrt{\epsilon}h^{00}+2\sqrt{\epsilon}\omega+\varphi. 
\label{full}\end{align}
The propagator of each mode can be expressed by that of a scalar $\phi$ with a certain mass:   
\begin{align}
\langle X(x)X(x')\rangle&=-\langle \phi(x)\phi(x')\rangle|_{m^2=0}, \notag\\
\langle \tilde{h}^{ij}(x)\tilde{h}^{kl}(x')\rangle&=(\delta^{ik}\delta^{jl}+\delta^{il}\delta^{jk}-\frac{2}{3}\delta^{ij}\delta^{kl})\langle \phi(x)\phi(x')\rangle|_{m^2=0}, \notag\\
\langle \Phi(x)\Phi(x')\rangle&=\langle \phi(x)\phi(x')\rangle|_{m^2=(3\eta-6\epsilon)H^2},
\label{compro}
\end{align}
\begin{align}
\langle Y(x)Y(x')\rangle&=\langle \phi(x)\phi(x')\rangle|_{m^2=2H^2}, \notag\\
\langle h^{0i}(x)h^{0j}(x')\rangle&=\delta^{ij}\langle \phi(x)\phi(x')\rangle|_{m^2=2H^2}. 
\end{align}

Since our interest is to evaluate inflationary correlation functions, we may neglect the modes whose mass-squares are $\mathcal{O}(1)H^2$. 
Eliminating them from (\ref{full}), we obtain the following identities: 
\begin{align}
h^{00}&=\frac{\sqrt{3}}{2}X-\frac{3}{2}\sqrt{\epsilon}\Phi, \notag\\
\omega&=\frac{\sqrt{3}}{4}X-\frac{1}{4}\sqrt{\epsilon}\Phi, \notag\\
\varphi&=\Phi. 
\label{light}\end{align}

$\tilde{h}_{ij}$, $h^{00}$, $\omega$ and $\varphi$ are left as light modes. 
$\tilde{h}_{ij}$ can be resolved to tensor, vector and scalar modes: 
\begin{align}
&\tilde{h}^{ij}=\tilde{h}^{ij}_T+\tilde{h}^{ij}_V+\tilde{h}^{ij}_S,\hspace{1em}\partial_i\tilde{h}^{ij}_T=0, \notag\\
&\tilde{h}^{ij}_V=\frac{\partial_i}{\sqrt{\partial_k^2}}V^j+\frac{\partial_j}{\sqrt{\partial_k^2}}V^i,\hspace{1em}\partial_iV^i=0, \notag\\
&\tilde{h}^{ij}_S=\sqrt{3}(\frac{\partial_i\partial_j}{\partial_k^2}-\frac{1}{3}\delta_{ij})S,  
\label{TVS}\end{align}
where $\tilde{h}^{ij}_T$ is gauge invariant. 
Another gauge invariant quantity at the linear perturbation level is given by 
\begin{align} 
\zeta&=\zeta^B-\frac{1}{\sqrt{2\epsilon}M_\text{pl}}\varphi,\hspace{1em}\zeta^B=\frac{\sqrt{2}}{M_\text{pl}}(\omega+\frac{1}{6}h^{00}-\frac{1}{2\sqrt{3}}S), 
\label{zetadef}
\end{align}
where $\zeta^B$ is the curvature perturbation in the BRST gauge and $\zeta$ corresponds to that in the comoving gauge. 

The free curvature perturbation is expanded as 
\begin{align}
\zeta=\int\frac{d^3k}{(2\pi)^3}\ \zeta_{\bf k} e^{i{\bf k}\cdot{\bf x}}, 
\end{align}
\begin{align}
\langle \zeta_{\bf k}\zeta_{{\bf k}'}\rangle=(2\pi)^3\delta^{(3)}({\bf k}+{\bf k}')\cdot 
\frac{H_*^2}{2k^3}\cdot\frac{1}{2\epsilon_* M_\text{pl}^2},
\end{align}
where $H$ and $\epsilon$ are evaluated on the horizon. 
The free graviton is expanded as 
\begin{align}
\tilde{h}_{ij}^T=\int\frac{d^3k}{(2\pi)^3}\ \sum_{s=\pm}\epsilon_{ij}^s({\bf k})T_{\bf k}^s e^{i{\bf k}\cdot{\bf x}},
\end{align}
\begin{align}
\langle T_{\bf k}^s T_{{\bf k}'}^{s'}\rangle=(2\pi)^3\delta^{(3)}({\bf k}+{\bf k}')\cdot\delta^{ss'}\cdot\frac{H_*^2}{2k^3}, 
\end{align}
where the polarization tensor satisfies the following identities: 
\begin{align}
k_i\epsilon_{ij}^s=\epsilon_{ii}^s=0, 
\end{align} 
\begin{align}
\epsilon_{ij}^s({\bf k})\epsilon_{kl}^s({\bf k})=(\bar{\delta}_{ik}\bar{\delta}_{jl}+\bar{\delta}_{il}\bar{\delta}_{jk}-\bar{\delta}_{ij}\bar{\delta}_{kl}),\hspace{1em}\bar{\delta}_{ij}=\delta_{ij}-\frac{k_ik_j}{k^2}, 
\end{align}
\begin{align}
\epsilon_{ij}^s({\bf k})\epsilon_{ij}^{s'}({\bf k})=2\delta^{ss'}. 
\end{align}

In order to respect the BRST symmetry, we need to introduce the Faddeev--Popov ghost term: 
\begin{align}
\mathcal{L}_\text{FP}&=-\sqrt{-\hat{g}}\hat{g}^{\mu\nu}\bar{b}_\nu\delta F_\mu|_{\delta x^\mu\to b^\mu} \notag\\
&=-\sqrt{-\hat{g}}\hat{g}^{\mu\nu}\bar{b}_\nu \{\delta_\mu^{\ \rho}\partial_\sigma-\frac{1}{2}\delta_\sigma^{\ \rho}\partial_\mu+2\delta_\sigma^{\ 0}\delta_\mu^{\ \rho}a^{-1}\partial_0a\}\delta\bar{h}_\rho^{\ \sigma}|_{\delta x^\mu\to b^\mu} \notag\\
&=a^2\partial_\nu\bar{b}^\mu\delta \bar{h}_\mu^{\ \nu}-\frac{1}{2}a^2\partial_\mu \bar{b}^\mu\delta \bar{h}_\nu^{\ \nu}-a\partial_0a\bar{b}^0\delta \bar{h}_\nu^{\ \nu}|_{\delta x^\mu\to b^\mu}, 
\label{FP}\end{align}
where $b^\mu$ is the ghost field, $\bar{b}^\mu$ is the anti-ghost field and $\bar{h}_\mu^{\ \nu}\equiv h_\mu^{\ \nu}+2\omega\delta_\mu^{\ \nu}$. 
Up to the leading order of the slow-roll parameters, the quadratic Faddeev--Popov ghost term is written as 
\begin{align}
\mathcal{L}_\text{FP}|_2&=-a^2\partial_\mu \bar{b}_\nu\partial^\mu b^\nu-(6\partial_0a\partial_0a-2a\partial_0^2a)\bar{b}_0b^0 \notag\\
&=-a^2\partial_\mu \bar{b}^i\partial^\mu b^i+a^2\partial_\mu \bar{b}^0\partial^\mu b^0+2H^2a^4\bar{b}^0b^0, 
\label{FP2}\end{align}
and then the corresponding propagators are given by 
\begin{align}
\langle b^i(x)\bar{b}^j(x')\rangle=\delta^{ij}\langle \phi(x)\phi(x')\rangle|_{m^2=0}, 
\end{align}
\begin{align}
\langle b^0(x)\bar{b}^0(x')\rangle=-\langle \phi(x)\phi(x')\rangle|_{m^2=2H^2}. 
\end{align}

\section{Non-Gaussianity}
\setcounter{equation}{0}

In this section, we identify the scalar and tensor perturbations in the BRST gauge with those in the comoving gauge
which become constant outside the cosmological horizon. 
It is realized by a gauge transformation of the variables in the BRST gauge into those
of the comoving gauge form.
Our construction reproduces the identical non-Gaussianity with the standard comoving gauge.

As the first example, we evaluate $\langle \zeta_{{\bf k}_1}\zeta_{{\bf k}_2}\zeta_{{\bf k}_3}\rangle$.   
The relevant vertices are given by  
\begin{align}
S_{\zeta\zeta\zeta}&=\frac{1}{2}\frac{\sqrt{2}}{M_\text{pl}}\int d^4x\ \big[-2a^2\omega\partial_\mu\varphi\partial^\mu\varphi+a^2h^{00}\partial_0\varphi\partial_0\varphi+\frac{1}{3}a^2h^{00}\partial_i\varphi\partial_i\varphi \notag\\
&\hspace{7.5em}+\sqrt{3}\big(\frac{\partial_i\partial_j}{\partial_k^2}-\frac{1}{3}\delta_{ij}\big)a^2S\partial_i\varphi\partial_j\varphi\big]. 
\label{zzz}\end{align}
Up to the second order, the BRST gauge is translated to the comoving gauge as 
\begin{align}
\zeta=\zeta^B -\sqrt{\frac{3}{2}}\partial_k(-\frac{1}{\sqrt{2\epsilon}}\frac{\varphi}{M_\text{pl}})\frac{\partial_k}{\partial_l^2}\frac{S}{M_\text{pl}}+\frac{\epsilon-\frac{1}{2}\eta}{2\epsilon}\frac{\varphi^2}{M_\text{pl}^2}, 
\label{translation1}\end{align}
\begin{align}
\gamma_{ij}=\gamma_{ij}^B-\sqrt{\frac{3}{2}}\partial_k\big(\frac{\sqrt{2}}{M_\text{pl}}\tilde{h}_{ij}^T\big)\frac{\partial_k}{\partial_l^2}\frac{S}{M_\text{pl}},\hspace{1em}\gamma_{ij}^B=\frac{\sqrt{2}}{M_\text{pl}}\tilde{h}_{ij}^T. 
\label{translation2}\end{align}
We show how to derive the gauge translation in Appendix \ref{B}. 

As seen in (\ref{light}), the soft graviton and soft inflaton are correlated in a slow-roll inflation: $\langle h^{00}\varphi\rangle,\langle \omega\varphi\rangle=\mathcal{O}(\epsilon)$. 
It should be noted that in evaluating the non-Gaussianity, we cannot neglect these cross correlations. 
From (\ref{light}), (\ref{zzz}) and (\ref{translation1}), $\langle \zeta_{{\bf k}_1}\zeta_{{\bf k}_2}\zeta_{{\bf k}_3}\rangle$ is evaluated as
\begin{align}
\langle \zeta_{{\bf k}_1}\zeta_{{\bf k}_2}\zeta_{{\bf k}_3}\rangle 
=&\ (2\pi)^3\delta^{(3)}({\bf k}_1+{\bf k}_2+{\bf k}_3)\cdot \frac{H^6}{(2k_1^3)(2k_2^3)(2k_3^3)} \notag\\
&\times\frac{i}{2\epsilon M_\text{pl}^4}\big\{-2I_1({\bf k}_1;{\bf k}_2,{\bf k}_3)+2I_1({\bf k}_1;{\bf k}_2,{\bf k}_3)+\frac{2}{3}I_2({\bf k}_1;{\bf k}_2,{\bf k}_3)-I_3({\bf k}_1;{\bf k}_2,{\bf k}_3) \notag\\
&\hspace{4.5em}+\text{(perms.)}\big\} \notag\\
&+(2\pi)^3\delta^{(3)}({\bf k}_1+{\bf k}_2+{\bf k}_3)\cdot \frac{H^6}{(2k_1^3)(2k_2^3)(2k_3^3)} \notag\\
&\times\frac{1}{2\epsilon M_\text{pl}^4}\big\{\frac{({\bf k}_1\cdot{\bf k}_2)k_3^3+({\bf k}_1\cdot{\bf k}_3)k_2^3}{H^2k_1^2} +\text{(perms.)}\big\} \notag\\
&+(2\pi)^3\delta^{(3)}({\bf k}_1+{\bf k}_2+{\bf k}_3)\cdot \frac{H^6}{(2k_1^3)(2k_2^3)(2k_3^3)}\cdot\frac{(\epsilon-\frac{1}{2}\eta)(k_1^3+k_2^3+k_3^3)}{\epsilon^2 M_\text{pl}^4 H^2}. 
\end{align}
where $I_1$, $I_2$ and $I_3$ are defined as 
\begin{align}
I_1({\bf k}_1;{\bf k}_2,{\bf k}_3) 
=&\ \big[\int^{\tau}_{-\infty}\frac{d\tau'}{(-H\tau')^2}\ (1-ik_1\tau')k_2^2\tau'k_3^2\tau'e^{i(k_1+k_2+k_3)\tau'}-\text{(h.c.)}\big] \notag\\
=&\ \frac{k_2^2k_3^2}{H^2}\cdot -2i\big\{\frac{1}{k_1+k_2+k_3}+\frac{k_1}{(k_1+k_2+k_3)^2}\big\}, 
\label{I_1}\end{align}
\begin{align}
I_2({\bf k}_1;{\bf k}_2,{\bf k}_3) 
=&-({\bf k}_2\cdot{\bf k}_3) \label{I_2}\\
&\times\big[\int^{\tau}_{-\infty}\frac{d\tau'}{(-H\tau')^2}\ (1-ik_1\tau')(1-ik_2\tau')(1-ik_3\tau')e^{i(k_1+k_2+k_3)\tau'} -\text{(h.c.)}\big] \notag\\
=&-\frac{({\bf k}_2\cdot{\bf k}_3)}{H^2}\cdot2i\big\{-(k_1+k_2+k_3)+\frac{k_1k_2+k_2k_3+k_3k_1}{k_1+k_2+k_3}+\frac{k_1k_2k_3}{(k_1+k_2+k_3)^2}\big\}, \notag
\end{align}
\begin{align}
I_3({\bf k}_1;{\bf k}_2,{\bf k}_3) 
=&-\big(\frac{({\bf k}_1\cdot{\bf k}_2)({\bf k}_1\cdot{\bf k}_3)}{k_1^2}-\frac{1}{3}({\bf k}_2\cdot{\bf k}_3)\big) \notag\\
&\times\big[\int^{\tau}_{-\infty}\frac{d\tau'}{(-H\tau')^2}\ (1-ik_1\tau')(1-ik_2\tau')(1-ik_3\tau')e^{i(k_1+k_2+k_3)\tau'}-\text{(h.c.)}\big] \notag\\
=&-\frac{({\bf k}_1\cdot{\bf k}_2)({\bf k}_1\cdot{\bf k}_3)/k_1^2-\frac{1}{3}({\bf k}_2\cdot{\bf k}_3)}{H^2} \notag\\
&\times 2i\big\{-(k_1+k_2+k_3)+\frac{k_1k_2+k_2k_3+k_3k_1}{k_1+k_2+k_3}+\frac{k_1k_2k_3}{(k_1+k_2+k_3)^2}\big\}. 
\label{I_3}\end{align}

After cumbersome calculations, we obtain
\begin{align}
&\ i\big\{\frac{2}{3}I_2({\bf k}_1;{\bf k}_2,{\bf k}_3)-I_3({\bf k}_1;{\bf k}_2,{\bf k}_3)\big\}+\frac{({\bf k}_1\cdot{\bf k}_2)k_3^3+({\bf k}_1\cdot{\bf k}_3)k_2^3}{H^2k_1^2} \notag\\
=&\ \frac{1}{H^2}\big\{-\frac{1}{2}k_1^3+\frac{1}{2}k_1(k_2^2+k_3^2)+\frac{4k_2^2k_3^2}{k_1+k_2+k_3}\big\}, 
\label{cumbersome1}\end{align}
and then $\langle \zeta_{{\bf k}_1}\zeta_{{\bf k}_2}\zeta_{{\bf k}_3}\rangle$ is written as 
\begin{align}
\langle \zeta_{{\bf k}_1}\zeta_{{\bf k}_2}\zeta_{{\bf k}_3}\rangle 
=&\ (2\pi)^3\delta^{(3)}({\bf k}_1+{\bf k}_2+{\bf k}_3)\cdot \frac{H^6}{(2k_1^3)(2k_2^3)(2k_3^3)} \label{result1}\\
&\times\frac{1}{H^2 M_\text{pl}^4}\Big[\frac{1}{2\epsilon}\big\{-\frac{1}{2}\sum_i k_i^3+\frac{1}{2}\sum_{i\not=j}k_ik_j^2+\frac{4\sum_{i<j}k_i^2k_j^2}{\sum_k k_k}\big\}+\frac{\epsilon-\frac{1}{2}\eta}{\epsilon^2}\sum_i k_i^3\Big]. \notag
\end{align}


In a similar way, we can evaluate $\langle \zeta_{{\bf k}_1}\gamma_{{\bf k}_2}^{s_2}\gamma_{{\bf k}_3}^{s_3}\rangle$. 
The relevant vertices are given by 
\begin{align}
S_{\zeta\gamma\gamma}
&=\frac{\sqrt{2}}{M_\text{pl}}\int d^4x\ \big[-\frac{1}{2}a^2\omega\partial_\mu \tilde{h}_{ij}^T\partial^\mu \tilde{h}_{ij}^T +\frac{1}{4}a^2h^{00}\partial_0 \tilde{h}_{ij}^T\partial_0 \tilde{h}_{ij}^T +\frac{1}{12}a^2h^{00}\partial_k \tilde{h}_{ij}^T\partial_k \tilde{h}_{ij}^T \notag\\
&\hspace{6.7em}+\frac{\sqrt{3}}{4}(\frac{\partial_k\partial_l}{\partial_m^2}-\frac{1}{3}\delta_{kl})a^2S\partial_k \tilde{h}_{ij}^T\partial_l \tilde{h}_{ij}^T\big], 
\label{zgg}\end{align} 
and then $\langle \zeta_{{\bf k}_1}\gamma_{{\bf k}_2}^{s_2}\gamma_{{\bf k}_3}^{s_3}\rangle$ is evaluated as 
\begin{align}
\langle \zeta_{{\bf k}_1}\gamma_{{\bf k}_2}^{s_2}\gamma_{{\bf k}_3}^{s_3}\rangle 
=&\ (2\pi)^3\delta^{(3)}({\bf k}_1+{\bf k}_2+{\bf k}_3)\cdot \epsilon_{ij}^{s_2}({\bf k}_2)\epsilon_{ij}^{s_3}({\bf k}_3)\cdot\frac{H^6}{(2k_1^3)(2k_2^3)(2k_3^3)} \notag\\
&\times\frac{i}{M_\text{pl}^4}\big\{\frac{2}{3}I_2({\bf k}_1;{\bf k}_2,{\bf k}_3)-I_3({\bf k}_1;{\bf k}_2,{\bf k}_3)\big\} \notag\\
&+(2\pi)^3\delta^{(3)}({\bf k}_1+{\bf k}_2+{\bf k}_3)\cdot \epsilon_{ij}^{s_2}({\bf k}_2)\epsilon_{ij}^{s_3}({\bf k}_3)\cdot\frac{H^6}{(2k_1^3)(2k_2^3)(2k_3^3)} \notag\\
&\times\frac{1}{M_\text{pl}^4}\big\{\frac{({\bf k}_1\cdot{\bf k}_2)k_3^3+({\bf k}_1\cdot{\bf k}_3)k_2^3}{H^2k_1^2}\big\} \notag\\
=&\ (2\pi)^3\delta^{(3)}({\bf k}_1+{\bf k}_2+{\bf k}_3)\cdot \epsilon_{ij}^{s_2}({\bf k}_2)\epsilon_{ij}^{s_3}({\bf k}_3)\cdot\frac{H^6}{(2k_1^3)(2k_2^3)(2k_3^3)} \notag\\
&\times\frac{1}{H^2M_\text{pl}^4}\big\{-\frac{1}{2}k_1^3+\frac{1}{2}k_1(k_2^2+k_3^2)+\frac{4k_2^2k_3^2}{k_1+k_2+k_3}\big\}. 
\label{result2}\end{align}


Comparing (\ref{result1}) and (\ref{result2}) with (\ref{result1'}) and (\ref{result2'}),  
we find that the computation results in BRST formalism are consistent with those in the comoving gauge. 

\section{Infrared logarithmic effects  to scalar perturbation}
\setcounter{equation}{0}

The accumulation of almost scale invariant fluctuations could give rise to logarithmic corrections to sub-horizon dynamics at the loop level. 
We investigate IR logarithmic effects to scalar perturbation in a single field inflation theory. 
The inflaton field is decomposed as the classical and quantum parts: $\hat{\varphi}(x)+\varphi (x)$. 
Since we focus on a local evolution of an inflaton field, we may assume $\hat{\varphi}(x)\ll M_\text{pl}$. 
We expand the potential term as 
\begin{align}
V(\hat{\varphi}(x)+\varphi (x))=V(\hat{\varphi}(x))+V'(\hat{\varphi}(x))\varphi (x)+\frac{1}{2}V''(\hat{\varphi}(x))\varphi^2 (x)+\cdots, 
\end{align} 
where we may neglect higher-other terms in slow-roll as each derivative of the potential
introduces a suppression factor $O(\sqrt{\epsilon})$.
We conformally rescale the inflaton field as
\begin{align}
\hat{\varphi}(x)+\varphi (x) =e^{-\kappa \omega(x)}(\hat{\varphi} (x)+\tilde{\varphi}(x)). 
\end{align} 
At the tree level, this is nothing but a change of variables, and $\varphi (x)$ can be expressed by $\tilde{\varphi}(x)$ through this relation.
However, it is $\tilde{\varphi}(x)$ which respects Lorentz symmetry at the sub-horizon scale with respect to the IR logarithmic corrections \cite{KitamotoSD}. 
To the leading order of the slow-roll parameter, the scalar perturbation is 
\begin{align}
\frac{\kappa^2}{4\epsilon}\langle\varphi(x)\varphi (x)\rangle=\frac{\kappa^2}{4\epsilon}\langle(\tilde{\varphi}(x)-\kappa \omega(x)\hat{\varphi} (x))^2\rangle. 
\end{align} 
As we consider the local evolution of the inflaton field $\hat{\varphi}(x)\ll M_\text{pl}$, we can neglect the $\kappa \omega(x)\hat{\varphi} (x)$ term. 

We have examined the one-loop infrared logarithmic effect to the kinetic term of $\tilde{\varphi}(x)$, 
\begin{align}
-\int d^4x\ \frac{1}{2}a^2(\tau)\big\{1+\frac{1}{2}\frac{\kappa^2H^2}{4\pi^2}\log a(\tau)\big\}\partial_{\mu}\tilde{\varphi}\partial^{\mu}\tilde{\varphi}. 
\end{align} 
The $\frac{1}{4}$ of this effect comes from the inflaton super-horizon mode. 
This effect can be canceled by the wave function renormalization of $\tilde{\varphi}\to Z\tilde{\varphi}$, where $Z=\big\{1-\frac{1}{4}\frac{\kappa^2H^2}{4\pi^2}\log a(\tau)\big\}$ at the sub-horizon scale. 
We may canonically normalize the $\tilde{\varphi}$ field by using this freedom of the wave function renormalization.\footnote{We obtain a finite mass correction $\delta m^2=-{\kappa^2H^4/16\pi^2}$ after this procedure. 
There is a subtraction ambiguity with respect to finite corrections, which may be fixed by shift symmetry.} 

However, we also need to consider IR logarithmic corrections  to the slow-roll parameter $\epsilon$.
We recall the scalar perturbation is given by
\begin{align}
\zeta=-H\delta t = -H  \frac{\varphi}{\dot{\hat{\varphi}}}=-\frac{\kappa}{2}\frac{\varphi}{\sqrt{\epsilon}}, 
\end{align} 
\begin{align}
2k^3\langle\zeta_{\bf k}\zeta_{-{\bf k}}\rangle'\sim \frac{1}{4\epsilon}\kappa^2H^2, 
\end{align} 
where $\langle\hspace{1em}\rangle'$ means that $(2\pi)^3\delta^{(3)}({\bf 0})$ is omitted. 
The slow-roll parameter $\epsilon$ is
\begin{align}
\epsilon = \frac{\kappa^2}{4}\big(\frac{\dot{\hat{\varphi}}}{H}\big)^2= \frac{1}{\kappa^2}\big(\frac{V'(\hat{\varphi})}{V(\hat{\varphi})}\big)^2. 
\label{defeps}
\end{align} 
We thus estimate IR logarithmic effects to $V'(\hat{\varphi}),V(\hat{\varphi})$ and $\kappa^2$.

\begin{figure}[t]
\begin{center}
\includegraphics[clip,width=12cm]{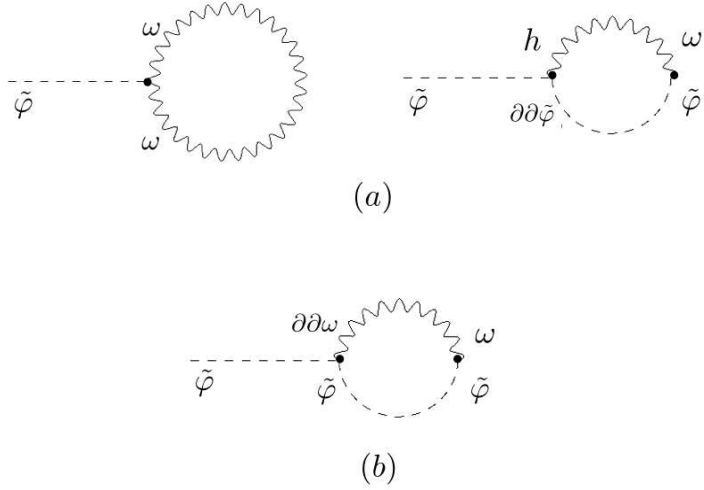}
\caption{These one-loop diagrams contribute IR logarithmic correction to the one-point operator. 
The broken line denotes the scalar propagator and the wavy line denotes the graviton propagator. 
The diagrams in $(a)$ are from soft gravitational fluctuation, and $(b)$ is from the soft inflaton. 
The propagator which contains derivatives has no IR logarithmic correction. }
\label{fig:1pt}
\end{center}
\end{figure}

For this purpose, we first investigate the following one-point operator  containing $V'(\hat{\varphi})$: 
\begin{align}
\int d^4x\ a^4(\tau) V'(\hat{\varphi})e^{3\kappa \omega}Z\tilde{\varphi}. 
\label{V'}\end{align}
We consider the expectation values of the operators with respect to the Birrell-Davies vacuum $|0\rangle$. 
There are two contributions to the one-loop IR logarithmic correction in this one-point operator, from soft gravitational fluctuation and the soft inflaton. 
Diagrams which contain IR logarithmic corrections to this one-point operator are shown in Fig. \ref{fig:1pt}.\footnote{When we consider IR logarithmic corrections from the soft inflaton, we treat the graviton as a hard graviton. 
That is, we treat it as if it were in flat spacetime. So there are no correlations between $\omega$ and $h$.}
From these diagrams, we can calculate the IR logarithmic correction from soft gravitational fluctuation as
\begin{align}
V'(\hat{\varphi})\tilde{\varphi}\big\{1+\frac{9}{32}\frac{\kappa^2H^2}{4\pi^2}\log a(\tau)\big\}, 
\end{align}
and  that from the soft inflaton as
\begin{align}
V'(\hat{\varphi})\tilde{\varphi}\big\{1-\frac{3}{8}\frac{\kappa^2H^2}{4\pi^2}\log a(\tau)\big\}.
\end{align} 

From these results, we can calculate the one-loop IR logarithmic correction in the one-point operator  as
\begin{align}
\int d^4x a^4(\tau) V'(\hat{\varphi})\tilde{\varphi}\big\{1-\frac{11}{32}\frac{\kappa^2H^2}{4\pi^2}\log a(\tau)\big\}.
\end{align}

The IR logarithmic correction to the cosmological constant operator is
\begin{align}
\int d^4x\ a^4(\tau) V(\hat{\varphi})\big\{1-\frac{3}{2}\frac{\kappa^2H^2}{4\pi^2}\log a(\tau)\big\}, 
\label{eq:ccoperator}
\end{align}
while that to the gravitational coupling is \cite{review}
\begin{align}
\int d^4x\ a^4(\tau)\hat{R} \frac{1}{\kappa^2}\big\{1-\frac{3}{4}\frac{\kappa^2H^2}{4\pi^2}\log a(\tau)\big\}. 
\label{eq:coupling}
\end{align}
So, 
\begin{align}
\frac{1}{\kappa^2}\frac{1}{V(\hat{\varphi})^2} \big\{1+\frac{9}{4}\frac{\kappa^2H^2}{4\pi^2}\log a(\tau)\big\}. 
\end{align}

We conclude that $\epsilon$ acquires the following extra time dependence due to the IR logarithmic effect: 
\begin{align}
\big\{1+\frac{25}{16}\frac{\kappa^2H_*^2}{4\pi^2}\log a(\tau_*)\big\}\epsilon_*. 
\label{extra}
\end{align}
Here the scale factor $a(\tau_*)$ should be evaluated at the moment of the horizon exit of the relevant momentum just like $H_*, \epsilon_*$. 
Note that the dimensionless ratio $\kappa^2H^2$ does not receive IR logarithmic corrections \cite{review}. 
The scalar perturbation receives IR logarithmic corrections through that of $\epsilon$, while the tensor perturbation does not receive IR logarithmic corrections. 
The tensor to scalar ratio $16\epsilon$ acquires extra  time dependence as in (\ref{extra}) due to the IR logarithmic effect.\footnote{The IR logarithmic effect in the scalar perturbation $\langle\zeta\zeta\rangle$ remains the same 
even if we do not renormalize the $\tilde{\varphi}$ field. 
Since the kinetic term is no longer canonical for the inflaton, a $Z^4$ factor appears in the definition (\ref{defeps}): $\epsilon = ({\kappa^2/ 4})({\dot{\hat{\varphi}}/ H})^2=({Z^4/ \kappa^2})({V'(\hat{\varphi})/ V(\hat{\varphi})})^2$. 
There is no Z factor in (\ref{V'}) in this procedure. 
As the $\langle\tilde{\varphi}\tilde{\varphi}\rangle$ correlator produces a $Z^2$ factor, $\langle\zeta\zeta\rangle$ is proportional to $1/Z^2$.} 

\begin{figure}[t]
\begin{center}
\includegraphics[clip,width=12cm]{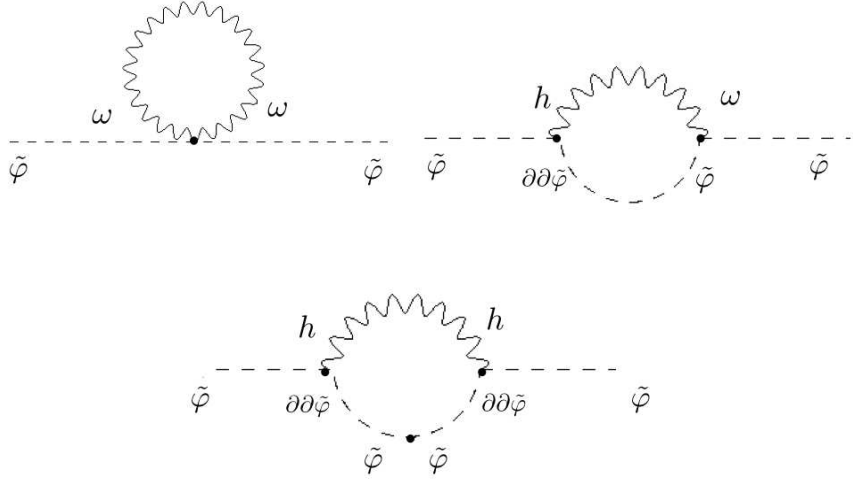}
\caption{The one-loop diagrams contribute an IR logarithmic correction from soft gravitational fluctuation to the two-point operator.}
\label{fig:2ptgraviton}
\end{center}
\end{figure}

\begin{figure}[t]
\begin{center}
\includegraphics[clip,width=12cm]{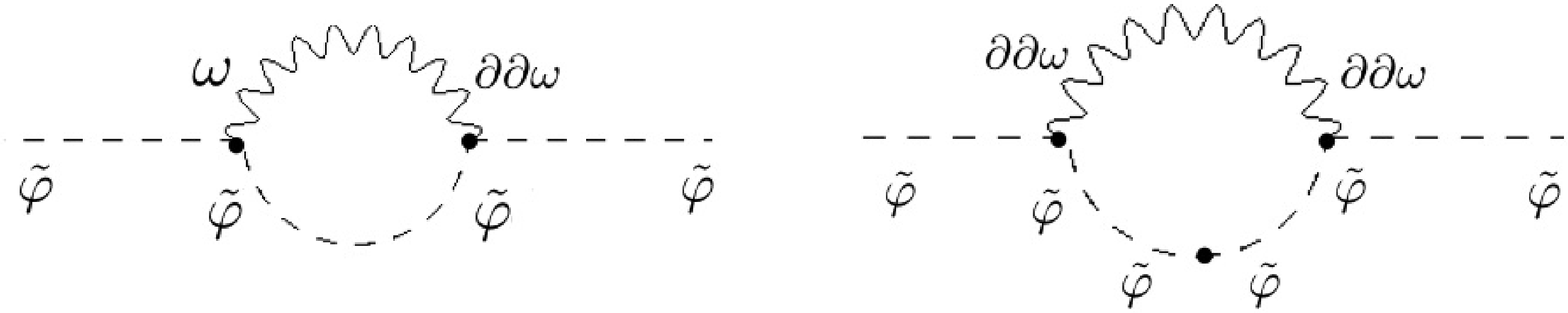}
\caption{The one-loop diagrams contribute an IR logarithmic correction from the soft inflaton to the two-point operator.}
\label{fig:2ptinflaton}
\end{center}
\end{figure}

We also consider IR logarithmic effects to another slow-roll parameter $\eta=\frac{2V''(\hat{\varphi})}{\kappa^2V(\hat{\varphi})}$. 
For this purpose, we investigate the following two-point operator containing $V''(\hat{\varphi})$: 
\begin{align}
\int d^4x\ a^4(\tau) V''(\hat{\varphi})e^{2\kappa \omega}Z^2\tilde{\varphi}^2.
\label{V''}\end{align}
Diagrams which contain IR logarithmic corrections from soft gravitational fluctuation to this operator are shown in Fig. \ref{fig:2ptgraviton}. 
We find that these IR logarithmic corrections from soft gravitational fluctuation are canceled. 
So, there are IR logarithmic corrections to this operator from the soft inflaton only. 
Diagrams which contain IR logarithmic corrections from the soft inflaton are shown in Fig. \ref{fig:2ptinflaton}.
From these diagrams, we can calculate the IR logarithmic correction from soft inflaton as
\begin{align}
V''(\hat{\varphi})\tilde{\varphi}^2\big\{1-\frac{1}{4}\frac{\kappa^2H^2}{4\pi^2}\log a(\tau)\big\}. 
\end{align}
So, the one-loop IR logarithmic correction in a two-point operator is
\begin{align}
\int d^4x\ a^4V''(\hat{\varphi})\tilde{\varphi}^2\big\{1-\frac{3}{4}\frac{\kappa^2H^2}{4\pi^2}\log a(\tau )\big\}.
\label{num}
\end{align}
From \eqref{eq:ccoperator} and \eqref{eq:coupling}, 
\begin{align}
\frac{1}{\kappa^2V(\hat{\varphi})}\big\{1+\frac{3}{4}\frac{\kappa^2H^2}{4\pi^2}\log a(\tau)\big\}. 
\label{den}
\end{align}
Because of the cancellation of (\ref{num}) and (\ref{den}), $\eta$ acquires no extra time dependence due to an IR logarithmic effect.\footnote{Alternatively, we may shift $\kappa \omega(x) \rightarrow \kappa (\omega(x) + \omega_c)$ 
to absorb the IR corrections to the cosmological and gravitational couplings. 
We still obtain the identical contributions to $\epsilon$ through the $e^{3\kappa \omega}$ factor in (\ref{V'}), and to $\eta$ through the $e^{2\kappa \omega}$ factor in (\ref{V''}). }


\section{Conclusion}
\setcounter{equation}{0}

We investigated cosmological correlation functions in a BRST gauge where locality and covariance at the sub-horizon scale are manifest. 
Since the cosmological perturbations are frozen at the super-horizon scale, it is determined by the dynamics of the sub-horizon scale. 
The slow-roll inflation is characterized by $\epsilon$ and $\eta$ parameters which represent the slope and curvature of the inflaton potential. 
Since they evolve with time even at the classical level, they could obtain IR logarithmic corrections. 
In fact we argue that $\epsilon$ receives IR logarithmic effects and undergoes additional quantum evolution. 
On the other hand, $\eta$ does not receive IR logarithmic correction. 
The covariance of the theory at the sub-horizon scale has played an important roll in determining IR logarithmic corrections to the slow-roll parameters. 
In this respect, the BRST gauge has the advantage of keeping manifest locality and covariance at the sub-horizon scale. 

Due to IR logarithmic effects (\ref{extra}), (\ref{classical}) is modified at the quantum level as 
\begin{align}
\frac{1}{2H}\frac{\dot{\epsilon}}{\epsilon}=-\eta+2\epsilon+\frac{25}{32}\frac{\kappa^2H^2}{4\pi^2}. 
\end{align}
Let us consider the case when $\epsilon$, $\eta$ are vanishingly small at the beginning due to the shift symmetry. 
IR logarithmic effects make $\epsilon$ grow with time 
\begin{align}
\epsilon\sim \epsilon_0e^{\frac{25}{16}\frac{\kappa^2H_0^2}{4\pi^2}H_0t}, 
\end{align}
while $\eta$ remains small. 
In this way we obtain an inflation model with the linear potential 
\begin{align}
V=\frac{6H_0^2}{\kappa^2}(1-\sqrt{\epsilon}\kappa\hat{\varphi}). 
\end{align}
We find it remarkable that the requirement that the shift symmetry be broken only by IR logarithmic effects singles out such an inflation model.\footnote{The linear potential may be generated by nonperturbative effects in string theory \cite{Silverstein,McAllister}. } 

When $\epsilon$ exceeds $\frac{\kappa^2H_0^2}{4\pi^2}$, we enter the classical region where 
\begin{align}
\frac{1}{2H}\frac{\dot{\epsilon}}{\epsilon}=2\epsilon. 
\end{align}
The Hubble parameter decreases as 
\begin{align}
H^2=H_0^2(1-3\epsilon_0H_0t)^\frac{2}{3}, 
\end{align}
while the $\epsilon$ parameter behaves as 
\begin{align}
\epsilon=\epsilon_0(1-3\epsilon_0H_0t)^{-\frac{4}{3}}. 
\end{align}
The number of e-foldings while $\epsilon$ grows from $\epsilon_i$ to $\epsilon_f$ is 
\begin{align}
N=\frac{1}{4\epsilon_i}-\frac{1}{4\epsilon_f}. 
\end{align}

Observationally, the magnitude of the CMB temperature fluctuation implies 
$\frac{\kappa^2H^2}{4\pi^2}\sim 10^{-8}\epsilon$. For an e-folding number $N\sim 50$, we find  $\epsilon\sim 1/200$.
In our scenario, $\epsilon$ is generated by IR quantum effects from nothing.
It thus starts the classical evolution characteristic of linear potential when
$\epsilon_i\sim\frac{\kappa^2H_0^2}{4\pi^2}$.
There must be an extended period of inflation with $\epsilon\sim 10^\frac{16}{3}\epsilon_i$, $H^2\sim 10^{-\frac{8}{3}}H_0^2$ before the seeds of CMB fluctuations exit the horizon. 
An inflation model with linear potential is currently under scrutiny \cite{Planck2015}. 
We must observe the tensor mode soon if IR logarithmic corrections are responsible for shift symmetry breaking. 

\section*{Acknowledgment}
This work is supported by Grant-in-Aid for Scientific Research (B) No. 26287044 and (C) No. 16K05336. 
We thank Chong-Sun Chu, Kazunori Kohri and Takahiro Tanaka for discussions. 

\appendix

\section{Quantization by solving constraints}\label{A}
\setcounter{equation}{0}

We make a brief review of the quantization by solving constraints \cite{Maldacena2002}. 
In the ADM formalism, we write the metric as 
\begin{align}
ds^2=-N^2dt^2+g_{ij}(dx^i+N^idt)(dx^j+N^jdt), 
\end{align}
and then the action is written as 
\begin{align}
S&=\int \sqrt{g^{(3)}}dtd^3x\ \frac{M_\text{pl}^2}{2}\big[NR^{(3)}+N^{-1}(E_i^{\ j}E_j^{\ i}-E_i^{\ i}E_j^{\ j})\big] \\
&+\int \sqrt{g^{(3)}}dtd^3x\ \frac{1}{2}\big[N^{-1}(\dot{\hat{\varphi}}+\dot{\varphi}-N^i\partial_i\varphi)(\dot{\hat{\varphi}}+\dot{\varphi}-N^j\partial_j\varphi)-N g^{ij}\partial_i\varphi\partial_j\varphi-2NV(\hat{\varphi}+\varphi)\big], \notag
\end{align}
where  
\begin{align}
E_{ij}=\frac{1}{2}(\dot{g}_{ij}-g_{jk}\nabla_i N^k-g_{ik}\nabla_j N^k),\hspace{1em}E^i_{\ j}=g^{ik}E_{kj}. 
\end{align}

We choose the spatially-flat gauge: 
\begin{align}
g_{ij}=a^2(e^\gamma)_{ij},\hspace{1em}\partial_i\gamma_{ij}=\gamma_{ii}=0. 
\end{align}
Since time derivatives of $N$ and $N^i$ do not appear in the action, these components are not dynamical. 
Differentiating the action with respect to $N^i$ and $N$, we obtain 
\begin{align}
\nabla_j\big\{N^{-1}(E_i^{\ j}-\delta_i^{\ j}E_k^{\ k})\big\}-\frac{1}{M_\text{pl}^2}N^{-1}\partial_i\varphi(\dot{\varphi}-N^j\partial_j\varphi)=0, 
\label{EQ1}\end{align}
\begin{align}
R^{(3)}&-N^{-2}(E_i^{\ j}E_j^{\ i}-E_i^{\ i}E_j^{\ j}) \notag\\
&-\frac{1}{M_\text{pl}^2}\big\{N^{-2}(\dot{\varphi}-N^i\partial_i\varphi)(\dot{\varphi}-N^j\partial_j\varphi)+g^{ij}\partial_i\varphi\partial_j\varphi+2V(\varphi)\big\}=0. 
\label{EQ2}\end{align}
Solving the constraints, we can express $N$ and $N^i$ by dynamical variables.   

\subsection{Quadratic terms}

Up to the linear order, the solutions of (\ref{EQ1}) and (\ref{EQ2}) are given by 
\begin{align}
\delta N=N-1=\sqrt{\frac{\epsilon}{2}}\frac{\varphi}{M_\text{pl}}, 
\label{N1}\end{align}
\begin{align}
N^i=-\sqrt{\frac{\epsilon}{2}}\frac{\partial_i}{\partial_j^2}\frac{\dot{\varphi}}{M_\text{pl}}. 
\label{Ni1}\end{align}
From (\ref{N1}) and (\ref{Ni1}), the quadratic term of the scalar is given by  
\begin{align}
S_{\varphi\varphi}=\frac{M_\text{pl}^2}{2}\int dtd^3x\ \big[a^3\frac{\dot{\varphi}^2}{M_\text{pl}^2}-a\frac{\partial_i\varphi\partial_i\varphi}{M_\text{pl}^2}-(3\eta-6\epsilon)H^2a^3\frac{\varphi^2}{M_\text{pl}^2}\big]. 
\end{align}
The spectrum of the scalar is not frozen at super-horizon scales except for $\dot{\epsilon}=0\Leftrightarrow \eta=2\epsilon$. 

At the linear order, the spatially-flat gauge is translated to the comoving gauge as 
\begin{align}
\zeta=-\frac{1}{\sqrt{2\epsilon}}\frac{\varphi}{M_\text{pl}}. 
\end{align}
The quadratic term of the curvature perturbation is written as 
\begin{align}
S_{\zeta\zeta}=\frac{M_\text{pl}^2}{2}\int dtd^3x\ \big[2\epsilon a^3\dot{\zeta}\dot{\zeta}-2\epsilon a\partial_i\zeta\partial_i\zeta\big], 
\end{align}
and so the spectrum of the curvature perturbation is frozen at super-horizon scales. 

In a similar way, we obtain the quadratic term of the graviton: 
\begin{align}
S_{\gamma\gamma}=\frac{M_\text{pl}^2}{2}\int dtd^3x\ \big[\frac{1}{4}a^3\dot{\gamma}_{ij}\dot{\gamma}_{ij}-\frac{1}{4}a\partial_k\gamma_{ij}\partial_k\gamma_{ij}\big]. 
\end{align}
The spectrum of graviton is frozen at super-horizon scales as well as that of the curvature perturbation. 

\subsection{Cubic terms}

Up to the second order, the solutions of (\ref{EQ1}) and (\ref{EQ2}) are given by 
\begin{align}
\delta N=\sqrt{\frac{\epsilon}{2}}\frac{\varphi}{M_\text{pl}}+\frac{1}{8H}\frac{\partial_i}{\partial_l^2}(\dot{\gamma}_{jk}\partial_i\gamma_{jk})+\frac{1}{2H}\frac{\partial_i}{\partial_j^2}(\frac{\dot{\varphi}}{M_\text{pl}}\partial_i\frac{\varphi}{M_\text{pl}}), 
\label{N2}\end{align}
\begin{align}
N^i=-\sqrt{\frac{\epsilon}{2}}\frac{\partial_i}{\partial_j^2}\frac{\dot{\varphi}}{M_\text{pl}}
-\frac{1}{4H}\frac{\partial_i}{\partial_j^2}\big\{&\ \frac{1}{4}\dot{\gamma}_{kl}\dot{\gamma}_{kl}+\frac{1}{4}a^{-2}\partial_m\gamma_{kl}\partial_m\gamma_{kl}+\frac{3H}{2}\frac{\partial_m}{\partial_n^2}(\dot{\gamma}_{kl}\partial_m\gamma_{kl}) \label{Ni2}\\
&+\frac{\dot{\varphi}^2}{M_\text{pl}^2}+a^{-2}\partial_k\frac{\varphi}{M_\text{pl}}\partial_k\frac{\varphi}{M_\text{pl}}+6H\frac{\partial_k}{\partial_l^2}(\frac{\dot{\varphi}}{M_\text{pl}}\partial_k\frac{\varphi}{M_\text{pl}})+3\eta H^2\frac{\varphi^2}{M_\text{pl}^2}\big\}. \notag
\end{align}

From (\ref{N2}) and (\ref{Ni2}), the cubic term for three scalars is given by 
\begin{align}
S_{\varphi\varphi\varphi}=\frac{M_\text{pl}^2}{2}\int a^3dtd^3x\ 
\big[&-\sqrt{\frac{\epsilon}{2}}\frac{\varphi}{M_\text{pl}}\frac{\dot{\varphi}^2}{M_\text{pl}^2}-\sqrt{\frac{\epsilon}{2}}a^{-2}\frac{\varphi}{M_\text{pl}}\frac{\partial_i\varphi\partial_i\varphi}{M_\text{pl}^2} \notag\\
&+2\sqrt{\frac{\epsilon}{2}}\frac{\dot{\varphi}}{M_\text{pl}}\partial_i\frac{\varphi}{M_\text{pl}}\frac{\partial_i}{\partial_j^2}\frac{\dot{\varphi}}{M_\text{pl}}\big]. 
\label{zzz'}\end{align}
Up to the second order, the spatially-flat gauge is translated to the comoving gauge as 
\begin{align}
\zeta=-\frac{1}{\sqrt{2\epsilon}}\frac{\varphi}{M_\text{pl}}+\frac{\epsilon-\frac{1}{2}\eta}{2\epsilon}\frac{\varphi^2}{M_\text{pl}^2}. 
\label{translation'}\end{align}
We extracted the terms which are dominant at super-horizon scales. 

From (\ref{zzz'}) and (\ref{translation'}), $\langle \zeta_{{\bf k}_1}\zeta_{{\bf k}_2}\zeta_{{\bf k}_3}\rangle$ is evaluated as 
\begin{align}
\langle \zeta_{{\bf k}_1}\zeta_{{\bf k}_2}\zeta_{{\bf k}_3}\rangle 
=&\ (2\pi)^3\delta^{(3)}({\bf k}_1+{\bf k}_2+{\bf k}_3)\cdot \frac{H^6}{(2k_1^3)(2k_2^3)(2k_3^3)}\notag\\
&\times\frac{i}{2\epsilon M_\text{pl}^4}\big\{\frac{1}{2}I_1({\bf k}_1;{\bf k}_2,{\bf k}_3)+\frac{1}{2}I_2({\bf k}_1;{\bf k}_2,{\bf k}_3)-I_4({\bf k}_1;{\bf k}_2,{\bf k}_3) \notag\\
&\hspace{4.5em}+\text{(perms.)}\big\} \notag\\
&+(2\pi)^3\delta^{(3)}({\bf k}_1+{\bf k}_2+{\bf k}_3)\cdot \frac{H^6}{(2k_1^3)(2k_2^3)(2k_3^3)}\cdot\frac{(\epsilon-\frac{1}{2}\eta)(k_1^3+k_2^3+k_3^3)}{\epsilon^2 H^2M_\text{pl}^4}, 
\end{align}
where $I_1$ and $I_2$ are defined in the same way as (\ref{I_1}) and (\ref{I_2}), and $I_4$ is defined as 
\begin{align}
I_4({\bf k}_1;{\bf k}_2,{\bf k}_3) 
=&\ \frac{{\bf k}_1\cdot{\bf k}_2}{2k_1^2}\big[\int^{\tau}_{-\infty}\frac{d\tau'}{(-H\tau')^2}\ (1-ik_2\tau')k_1^2\tau'k_3^2\tau'e^{i(k_1+k_2+k_3)\tau'}-\text{(h.c.)}\big] \notag\\
&+\frac{{\bf k}_1\cdot{\bf k}_3}{2k_1^2}\big[\int^{\tau}_{-\infty}\frac{d\tau'}{(-H\tau')^2}\ (1-ik_3\tau')k_1^2\tau'k_2^2\tau'e^{i(k_1+k_2+k_3)\tau'}-\text{(h.c.)}\big] \notag\\
=&\ \frac{({\bf k}_1\cdot{\bf k}_2)k_3^2}{2H^2}\cdot -2i\big\{\frac{1}{k_1+k_2+k_3}+\frac{k_2}{(k_1+k_2+k_3)^2}\big\} \notag\\
&+\frac{({\bf k}_1\cdot{\bf k}_3)k_2^2}{2H^2}\cdot -2i\big\{\frac{1}{k_1+k_2+k_3}+\frac{k_3}{(k_1+k_2+k_3)^2}\big\}. 
\label{I_4}\end{align}

After cumbersome calculations, we obtain
\begin{align}
&\ i\big\{\frac{1}{2}I_1({\bf k}_1;{\bf k}_2,{\bf k}_3)+\frac{1}{2}I_2({\bf k}_1;{\bf k}_2,{\bf k}_3)-I_4({\bf k}_1;{\bf k}_2,{\bf k}_3)\big\} \notag\\
=&\ \frac{1}{H^2}\big\{-\frac{1}{2}k_1^3+\frac{1}{2}k_1(k_2^2+k_3^2)+\frac{4k_2^2k_3^2}{k_1+k_2+k_3}\big\}, 
\label{cumbersome2}\end{align}
and then $\langle \zeta_{{\bf k}_1}\zeta_{{\bf k}_2}\zeta_{{\bf k}_3}\rangle$ is written as 
\begin{align}
\langle \zeta_{{\bf k}_1}\zeta_{{\bf k}_2}\zeta_{{\bf k}_3}\rangle 
=&\ (2\pi)^3\delta^{(3)}({\bf k}_1+{\bf k}_2+{\bf k}_3)\cdot \frac{H^6}{(2k_1^3)(2k_2^3)(2k_3^3)} \label{result1'}\\
&\times\frac{1}{H^2 M_\text{pl}^4}\Big[\frac{1}{2\epsilon}\big\{-\frac{1}{2}\sum_i k_i^3+\frac{1}{2}\sum_{i\not=j}k_ik_j^2+\frac{4\sum_{i<j}k_i^2k_j^2}{\sum_k k_k}\big\}+\frac{\epsilon-\frac{1}{2}\eta}{\epsilon^2}\sum_i k_i^3\Big]. \notag
\end{align}

Especially at the squeezed limit: $k_1\ll k_2,k_3$, $\langle \zeta_{{\bf k}_1}\zeta_{{\bf k}_2}\zeta_{{\bf k}_3}\rangle$ is evaluated as 
\begin{align}
\langle \zeta_{{\bf k}_1}\zeta_{{\bf k}_2}\zeta_{{\bf k}_3}\rangle
\to (2\pi)^3\delta^{(3)}({\bf k}_1+{\bf k}_2+{\bf k}_3)\cdot \frac{H^6}{(2k_1^3)(2k_2^3)(2k_3^3)}\cdot\frac{(3\epsilon-\eta)k_2^3}{\epsilon^2 H^2M_\text{pl}^4}. 
\label{squeezed1'}\end{align}

In a similar way, we obtain the cubic term for a scalar and two gravitons: 
\begin{align}
S_{\varphi\gamma\gamma}=\frac{M_\text{pl}^2}{2}\int a^3dtd^3x\ 
\big[&-\frac{1}{4}\sqrt{\frac{\epsilon}{2}}\frac{\varphi}{M_\text{pl}}\dot{\gamma}_{ij}\dot{\gamma}_{ij}-\frac{1}{4}\sqrt{\frac{\epsilon}{2}}a^{-2}\frac{\varphi}{M_\text{pl}}\partial_k\gamma_{ij}\partial_k\gamma_{ij} \notag\\
&+\frac{1}{2}\sqrt{\frac{\epsilon}{2}}\dot{\gamma}_{ij}\partial_k\gamma_{ij}\frac{\partial_k}{\partial_l^2}\frac{\dot{\varphi}}{M_\text{pl}}\big]. 
\label{zgg'}\end{align}
From (\ref{zgg'}), (\ref{translation'}) and (\ref{cumbersome2}), $\langle \zeta_{{\bf k}_1}\gamma_{{\bf k}_2}^{s_2}\gamma_{{\bf k}_3}^{s_3}\rangle$ is written as 
\begin{align}
\langle \zeta_{{\bf k}_1}\gamma_{{\bf k}_2}^{s_2}\gamma_{{\bf k}_3}^{s_3}\rangle
=&\ (2\pi)^3\delta^{(3)}({\bf k}_1+{\bf k}_2+{\bf k}_3)\cdot \epsilon_{ij}^{s_2}({\bf k}_2)\epsilon_{ij}^{s_3}({\bf k}_3)\cdot\frac{H^6}{(2k_1^3)(2k_2^3)(2k_3^3)} \notag\\
&\times\frac{i}{M_\text{pl}^4}\big\{\frac{1}{2}I_1({\bf k}_1;{\bf k}_2,{\bf k}_3)+\frac{1}{2}I_2({\bf k}_1;{\bf k}_2,{\bf k}_3)-I_4({\bf k}_1;{\bf k}_2,{\bf k}_3)\big\} \notag\\
=&\ (2\pi)^3\delta^{(3)}({\bf k}_1+{\bf k}_2+{\bf k}_3)\cdot \epsilon_{ij}^{s_2}({\bf k}_2)\epsilon_{ij}^{s_3}({\bf k}_3)\cdot\frac{H^6}{(2k_1^3)(2k_2^3)(2k_3^3)} \notag\\
&\times\frac{1}{H^2M_\text{pl}^4}\big\{-\frac{1}{2}k_1^3+\frac{1}{2}k_1(k_2^2+k_3^2)+\frac{4k_2^2k_3^2}{k_1+k_2+k_3}\big\}. 
\label{result2'}\end{align}

At the squeezed limit: $k_1\ll k_2,k_3$, $\langle \zeta_{{\bf k}_1}\gamma_{{\bf k}_2}^{s_2}\gamma_{{\bf k}_3}^{s_3}\rangle$ is evaluated as 
\begin{align}
\langle \zeta_{{\bf k}_1}\gamma_{{\bf k}_2}^{s_2}\gamma_{{\bf k}_3}^{s_3}\rangle
\to (2\pi)^3\delta^{(3)}({\bf k}_1+{\bf k}_2+{\bf k}_3)\cdot 2\delta^{s_2s_3}\cdot\frac{H^6}{(2k_1^3)(2k_2^3)(2k_3^3)}\cdot\frac{2k_2^3}{H^2M_\text{pl}^4}. 
\label{squeezed2'}\end{align}
At the other squeezed limit: $k_2\ll k_1,k_3$, it is evaluated as 
\begin{align}
\langle \zeta_{{\bf k}_1}\gamma_{{\bf k}_2}^{s_2}\gamma_{{\bf k}_3}^{s_3}\rangle
\to (2\pi)^3\delta^{(3)}({\bf k}_1+{\bf k}_2+{\bf k}_3)\cdot \epsilon_{ij}^{s_2}({\bf k}_2)\epsilon_{ij}^{s_3}({\bf k}_3)\cdot\frac{H^6}{(2k_1^3)(2k_2^3)(2k_3^3)}\cdot 0. 
\label{squeezed3'}\end{align}

The computation results (\ref{squeezed1'}), (\ref{squeezed2'}) and (\ref{squeezed3'}) can also be derived by considering the scaling of the propagators as 
\begin{align}
\langle \zeta_{{\bf k}_1}\zeta_{{\bf k}_2}\zeta_{{\bf k}_3}\rangle
&\to -\langle \zeta_{{\bf k}_1}\zeta_{-{\bf k}_1}\rangle' k_2\frac{d}{dk_2}\langle\gamma_{{\bf k}_2}^{s_2}\gamma_{{\bf k}_3}^{s_3}\rangle\hspace{1em}\text{at $k_1\ll k_2,k_3$}\notag\\
&=2\epsilon\langle \zeta_{{\bf k}_1}\zeta_{-{\bf k}_1}\rangle'\langle\gamma_{{\bf k}_2}^{s_2}\gamma_{{\bf k}_3}^{s_3}\rangle \notag\\
&=(2\pi)^3\delta^{(3)}({\bf k}_2+{\bf k}_3)\cdot \delta^{s_2s_3}\cdot\frac{H^4}{(2k_1^3)(2k_2^3)}\cdot\frac{2}{M_\text{pl}^4}, 
\end{align}
\begin{align}
\langle \zeta_{{\bf k}_1}\gamma_{{\bf k}_2}^{s_2}\gamma_{{\bf k}_3}^{s_3}\rangle
&\to -\langle \zeta_{{\bf k}_1}\zeta_{-{\bf k}_1}\rangle' k_2\frac{d}{dk_2}\langle\gamma_{{\bf k}_2}^{s_2}\gamma_{{\bf k}_3}^{s_3}\rangle\hspace{1em}\text{at $k_1\ll k_2,k_3$}\notag\\
&=2\epsilon\langle \zeta_{{\bf k}_1}\zeta_{-{\bf k}_1}\rangle'\langle\gamma_{{\bf k}_2}^{s_2}\gamma_{{\bf k}_3}^{s_3}\rangle \notag\\
&=(2\pi)^3\delta^{(3)}({\bf k}_2+{\bf k}_3)\cdot \delta^{s_2s_3}\cdot\frac{H^4}{(2k_1^3)(2k_2^3)}\cdot\frac{2}{M_\text{pl}^4}, 
\end{align}
\begin{align}
\langle \zeta_{{\bf k}_1}\gamma_{{\bf k}_2}^{s_2}\gamma_{{\bf k}_3}^{s_3}\rangle
&\to -\langle \gamma_{{\bf k}_2}^{s_2}\gamma_{-{\bf k}_2}^{s_2}\rangle' \epsilon^{s_2}_{ij}(k_1)^i(k_1)^j\frac{d}{dk_1^2}\langle\zeta_{{\bf k}_1}\gamma_{{\bf k}_3}^{s_3}\rangle\hspace{1em}\text{at $k_2\ll k_1,k_3$}\notag\\
&=0.
\end{align}

\section{From BRST gauge to comoving gauge}\label{B}
\setcounter{equation}{0}

Here we show how to translate the variables in the BRST gauge to those in the comoving gauge, up to the second order. 

As seen in (\ref{TVS}), the spatial metric in the BRST gauge contains the following modes: 
\begin{align}
\frac{\partial_i}{\sqrt{\partial_k^2}}V^j+\frac{\partial_j}{\sqrt{\partial_k^2}}V^i+\sqrt{3}\frac{\partial_i\partial_j}{\partial_k^2}S, 
\end{align} 
while that in the comoving gauge does not. 
For the translation to the comoving gauge, we need to eliminate them by using the spatial reparametrization: $x'^i=x^i+\delta x^i$. 
Up to the first order, $\delta x^i$ is given by  
\begin{align}
&\delta x^i=v^i+\partial_i s,\hspace{1em}\partial_i v^i=0, \notag\\
&v^i=\frac{\sqrt{2}}{M_\text{pl}}\frac{1}{\sqrt{\partial_k^2}}V^i,\hspace{1em}s=\frac{\sqrt{3}}{\sqrt{2}M_\text{pl}}\frac{1}{\partial_k^2}S. 
\label{spatial}\end{align}

The temporary reparametrization: $t'=t+\delta t$ is fixed by eliminating the scalar fluctuation. 
Up to the second order, 
\begin{align}
0=\varphi-\dot{\hat{\varphi}}\delta t-\frac{1}{2}\ddot{\hat{\varphi}}\delta t^2, 
\end{align}
and then $\delta t$ is given by 
\begin{align}
\delta t=\frac{1}{\dot{\hat{\varphi}}}\varphi-\frac{1}{2}\frac{\ddot{\hat{\varphi}}}{\dot{\hat{\varphi}}^3}\varphi^2. 
\label{temporary}\end{align}

From (\ref{spatial}) and (\ref{temporary}), the curvature perturbation and the graviton in the BRST gauge are translated to those in the comoving gauge as 
\begin{align}
\zeta&\simeq \zeta^B-H\delta t+\frac{1}{2}\epsilon H^2\delta t^2-\partial_k(\zeta^B-H\delta t)\delta x^k \notag\\
&\simeq\zeta^B-\frac{1}{\sqrt{2\epsilon}}\frac{\varphi}{M_\text{pl}}-\sqrt{\frac{3}{2}}\partial_k(-\frac{1}{\sqrt{2\epsilon}}\frac{\varphi}{M_\text{pl}})\frac{\partial_k}{\partial_l^2}\frac{S}{M_\text{pl}}+\frac{\epsilon-\frac{1}{2}\eta}{2\epsilon}\frac{\varphi^2}{M_\text{pl}^2}, 
\end{align}
\begin{align}
\gamma_{ij}&\simeq\gamma_{ij}^B-\partial_k\gamma_{ij}^B\delta x^k \notag\\
&\simeq\gamma_{ij}^B-\sqrt{\frac{3}{2}}\partial_k\gamma_{ij}^B\frac{\partial_k}{\partial_l^2}\frac{S}{M_\text{pl}}, 
\label{translation2'}\end{align}
where $\zeta^B$ and $\gamma^B$ are the scalar and tensor modes in the BRST gauge:  
\begin{align}
\zeta^B=\frac{\sqrt{2}}{M_\text{pl}}(\omega+\frac{1}{6}h^{00}-\frac{1}{2\sqrt{3}}S),\hspace{1em}\gamma_{ij}^B=\frac{\sqrt{2}}{M_\text{pl}}\tilde{h}_{ij}^T. 
\end{align}
We wrote down the relevant terms to evaluate the three-point functions at tree-level. 
Strictly speaking, the second term in (\ref{translation2'}) has a projection operator for transverse and traceless modes.   
However, the operator is not necessary because non-transverse and traceless modes are automatically canceled in evaluating the three-point functions. 


\end{document}